\documentclass[reprint,aps,prd,onecolumn,notitlepage,showpacs,nofootinbib,preprintnumbers,superscriptaddress]{revtex4-1}
\usepackage{amsmath}
\usepackage{amsfonts}
\usepackage{amssymb}
\usepackage{latexsym}
\usepackage{enumerate}
\usepackage{color,xcolor}
\usepackage{graphicx}
\usepackage{bm}
\usepackage{epsfig}
\usepackage[english]{babel}
\usepackage{hyperref}
\usepackage{times}
\usepackage{comment}
\usepackage{epstopdf}

\def\cm{\mathrm{cm}}
\def\mpc{\mathrm{Mpc}}

\def\omp{\omega_p}
\def\ro{r_\mathrm{o}}

\def\rph{r_\mathrm{ph}}
\def\rmin{r_\mathrm{min}}
\def\Rsh{R_\mathrm{sh}}

\def\ash{\alpha_\mathrm{sh}}
\def\ma{\dot{M}_A}
\def\sm{M_{\odot}}

\begin{document}

\title{Gravitational effect of plasma particles on the shadow of Schwarzschild black holes}
\author{Qiang Li}
\author{Yanni Zhu}
\author{Towe Wang}
\email[Electronic address: ]{twang@phy.ecnu.edu.cn}
\affiliation{Department of Physics, East China Normal University,\\
Shanghai 200241, China\\ \vspace{0.2cm}}
\date{\today\\ \vspace{1cm}}
\begin{abstract}
Considering a Schwarzschild black hole surrounded by a fully ionized hydrogen plasma, we study the effect of the gravitational field of the plasma particles on the shadow. We take a formalism in which this effect is unified with the refractive effect of the plasma medium studied previously, but the two effects are characterized by two independent parameters. For semi-realistic values of parameters, we find their corrections to the shadow radius are both negligible, and the gravitational correction can overtake the refractive correction for active galactic nuclei of masses larger than $10^9M_{\odot}$. With unrealistically large values of parameters, we illustrate the two effects on the light trajectories and the intensity map.
\end{abstract}


\maketitle




\section{Introduction}\label{sect-intro}
Black holes are a miracle of gravity. Although their existence used to be controversial in the last century, they are now believed to play important roles in the life cycle of a massive star, in the center of a galaxy and in the theory of quantum gravity. During the recent decade, more evidence for black holes has been obtained \cite{TheLIGOScientific:2016wfe,Abuter:2018drb,Akiyama:2019cqa}, including the amazing shadow and ring of M87* imaged by the Event Horizon Telescope (EHT) at a wavelength of $1.3$ millimeters \cite{Akiyama:2019cqa}. In the center of M87, there is a supermassive spinning black hole embedded in a geometrically thick, optically thin accretion disk. The shadow and ring is formed by strong gravitational lensing of synchrotron emission from the hot plasma in the disk. In this scenario, the plasma plays the part of a light source and the central black hole works as a very strong lens.

In the future, hopefully images of higher resolution and of other black holes will be generated, then it will become necessary to study the formation of the shadow and ring in more details. For example, the success of the EHT \cite{Akiyama:2019cqa} relied on the condition that the accretion disk of M87* is optically thin at millimeter wavelengthes. In this way, the plasma in the disk is treated as a pure emission source, and the small opacity has negligible influence on the fuzzy image. However, for other black holes, in longer baselines \cite{Roelofs:2019nmh} or at smaller wavelengthes \cite{Falcke:1999pj}, the plasma cannot always be transparent, then it will become necessary to consider the scattering and absorption of photons by the plasma.

Besides opacity, the plasma has two other effects on the shadow and ring. First, as a refractive medium the plasma can change the deflection angle of light rays in the gravitational field of a black hole \cite{BisnovatyiKogan:2008yg,BisnovatyiKogan:2010ar,2013ApSS346513M,Er:2013efa}. Second, the energy and momentum of the plasma particles can modify the gravitational field outside the black hole. For a Schwarzschild black hole surrounded by a spherically symmetric plasma, the former refractive effect has been investigated by Ref. \cite{Perlick:2015vta}. In this paper, we aim to study the latter gravitational effect in a similar toy model. In our model, we will assume that the plasma is static and its density profile falls as $r^{-3/2}$ outside the event horizon. This model is far from realistic, but it enables us to assess the gravitational effect of plasma particles following the method introduced in Ref. \cite{Perlick:2015vta} and make a comparison with the refractive effect of the plasma medium.

The rest of the paper is organized as follows. In Sec. \ref{sect-pre}, we briefly review some useful formulae in Ref. \cite{Perlick:2015vta} and establish our convention of notations. In Sec. \ref{sect-toy}, we extend the toy model of Ref. \cite{Perlick:2015vta} by incorporating the plasma's gravity into the metric. The density profile is cut off inside a sphere of certain critical radius. By specifying the cutoff radius, we consider two concrete models:
\begin{enumerate}[(A)]
\item The plasma density profile is cut off at the event horizon;
\item The plasma density profile is cut off at the innermost stable circular orbit.
\end{enumerate}
Applying the general formulae to model A, influences of the plasma on black hole shadows are investigated in Sec. \ref{sect-modA}. In Sec. \ref{subsect-radA} we work out the corrections of refractive and gravitational effects to radii of the photon sphere and the shadow. Inserting observational data of some active galactic nuclei (AGNs) \cite{Firouzjaee:2019aij,Daly:2020sex}, we find these corrections are negligibly small, though the gravitational correction can overtake the refractive correction for AGNs more massive than about $10^9M_{\odot}$. In Sec. \ref{sect-intA}, we illustrate the influences of refractive and gravitational effects on the trajectories of light rays and the observed intensity image. This is done by assigning exaggerated values to model parameters. For model B, a parallel investigation is done in Sec. \ref{sect-modB}. In Sec. \ref{sect-dis} we summarize the our results and discuss the implications.

\section{Previous results and convention of notations}\label{sect-pre}
We are interested in the shadow of a Schwarzschild black hole surrounded by a spherically symmetric plasma. In this situation, the shadow seen by a distant observer has the shape of a circular disk. To determine the radius of the shadow, one should study the equations of motion of light rays outside the black hole. This has been investigated in Ref. \cite{Perlick:2015vta} by including the refractive effect of the plasma medium but neglecting the gravitational field of the plasma particles. In this paper, we will take the gravitational effect of the plasma particles into account. In the current section, as a preparation, we will collect some useful formulae from Ref. \cite{Perlick:2015vta} (see also Ref. \cite{Perlick:2021aok}) which hold for spherical spacetimes generally.

In general, a static spherical spacetime is described by a metric of the form
\begin{equation}\label{metric-sph}
ds^2=-A(r)c^{2}dt^{2}+B(r)dr^{2}+D(r)\left(d\theta^{2}+\sin^{2}\theta d\phi^{2}\right)
\end{equation}
in which $c$ represents the light velocity. In accordance with the symmetry, the electron number density $N(r)$ is a function of radius only. If we denote the charge and mass of the electron as $e$ and $m_e$ respectively, then the plasma frequency is given by
\begin{equation}\label{omega-p}
\omp(r)^2=\frac{4\pi e^2}{m_e}N(r)
\end{equation}
in Gaussian units. It is useful to introduce the function
\begin{equation}\label{h}
h(r)^2=\frac{D(r)}{A(r)}\left[1-A(r)\frac{\omp(r)^2}{\omega_0^2}\right]
\end{equation}
with a constant of motion $\omega_0$. On the equatorial plane, the orbit equation of light rays can be written as
\begin{equation}\label{orbitr}
\frac{dr}{d\phi}=\pm\sqrt{\frac{D(r)}{B(r)}\left[\frac{h(r)^2}{b^2}-1\right]}
\end{equation}
with the parameter $b$ being a constant of motion. In terms of $h(r)^2$, the radius of the outermost photon sphere $\rph$ is the largest root of the equation
\begin{equation}\label{rph}
\frac{d}{dr}h(r)^2=0.
\end{equation}
For an observer at a distance of $r$, the opening angle of the shadow is determined by
\begin{equation}\label{ash}
\sin^2\ash=\frac{h(\rph)^2}{h(r)^2},
\end{equation}
and the radius-squared of the shadow is accordingly
\begin{equation}\label{bsh}
\Rsh^2=r^2\sin^2\ash=\frac{r^2h(\rph)^2}{h(r)^2}.
\end{equation}
More generally, it can be verified that in the image seen at a distance of $r$, photons with the same value of $b$ form a circle of radius-squared
\begin{equation}\label{R}
R^2=r^2\sin^2\alpha=\frac{b^2r^2}{h(r)^2}.
\end{equation}
If we restrict to asymptotically flat spacetimes, $\omega_0$ can be interpreted as the photon frequency at infinity, $b$ can be named as an impact parameter, while $\Rsh$ will be identical to the impact parameter of the photon sphere.


Eqs. \eqref{rph} and \eqref{bsh} are key formulae for us to compute radii of the photon sphere and the shadow, while Eqs. \eqref{orbitr} and \eqref{R} are crucial for tracing the trajectory of light rays. We will assume $h(r)\geq0$ and $0\leq\alpha\leq\pi/2$ for practical reasons. To trace the trajectory of light rays, one should integrate the orbit equation \eqref{orbitr}, usually using a numerical method. For the convenience of numerical integration, we will introduce a new variable $u=r_g^{1/2}/r^{1/2}$.

Throughout this paper, we work in Gaussian units. The mass of the black hole will be denoted as $M$, and its mass accretion rate will be denoted as $\ma$. For brevity, we introduce the gravitational radius $r_g=2GM/c^2$ with $G$ being the Newtonian constant and two dimensionless parameters
\begin{align}
\label{beta}\beta&=\frac{e^2\ma}{m_em_p\omega_0^2cr_g^2}\approx1.0\times10^{-7}\times\left(\frac{\lambda_0}{0.1\cm}\right)^2\left(\frac{10^9\sm}{M}\right)^2\left(\frac{\ma}{0.1\sm\mathrm{yr}^{-1}}\right),\\
\label{gamma}\gamma&=\frac{4G\ma}{3c^3}\approx2.1\times10^{-14}\times\left(\frac{\ma}{0.1\sm\mathrm{yr}^{-1}}\right).
\end{align}
The wavelength of the photon is related to its frequency by $\lambda_0=2\pi c/\omega_0$.

\section{Toy models}\label{sect-toy}
It is well known that the metric of the Schwarzschild black hole
\begin{equation}\label{metric-Sch}
ds^2=-\left(1-\frac{r_g}{r}\right)c^{2}dt^{2}+\left(1-\frac{r_g}{r}\right)^{-1}dr^{2}+r^2\left(d\theta^{2}+\sin^{2}\theta d\phi^{2}\right)
\end{equation}
is a static spherically symmetric solution of the vacuum Einstein equations. In Ref. \cite{Perlick:2015vta}, the refractive effect of a cold plasma on the shadow of Schwarzschild black holes has been evaluated in a concrete model \cite{1972ApSS15153M}, where the plasma has a mass density
\begin{equation}\label{den}
\rho(r)=\left\{\begin{array}{ll}
0, & \mathrm{for}~r\leq r_c,\\
\frac{\ma}{4\pi cr_g^{1/2}r^{3/2}}, & \mathrm{for}~r>r_c.
\end{array}\right.
\end{equation}
Here we have introduced a critical or cutoff radius $r_c$ for clarity. For a fully ionized hydrogen plasma, the plasma mass density is related to the electron number density by $\rho(r)=m_pN(r)$.

The gravitational field of the plasma has been neglected in Ref. \cite{Perlick:2015vta}. If we take the plasma's gravity into consideration, the Schwarzschild metric will be modified into a time-dependent metric. The reasons are as follows. First, owing to the accretion, the mass or equivalently radius of the black hole must increase with time. Second, even if we ignore the accretion, the plasma cannot keep static against the gravitational attraction of the central black hole. Consequently, the modified metric ought to be time-dependent both inside and outside the event horizon. This will prevent us from applying the formulae shown in Sec. \ref{sect-pre}. In order to circumvent this obstacle, it is customary to neglect the time dependence and assume \cite{Xu:2018wow,Konoplya:2019sns}
\begin{equation}
A(r)=\frac{1}{B(r)},~~~~D(r)=r^2
\end{equation}
in Eq. \eqref{metric-sph}. Making use of one of the Einstein equations, one can verify that
\begin{align}
\frac{1}{B(r)}&=1-\frac{2Gm(r)}{c^2r},\\
m(r)&=\left\{\begin{array}{ll}
M, & \mathrm{for}~r\leq r_c,\\
M+4\pi\int_{r_c}^{r}\rho(r')r'^2dr', & \mathrm{for}~r>r_c.
\end{array}\right.
\end{align}
Working out the integral with the mass density \eqref{den}, we find
\begin{equation}\label{m}
m(r)=\left\{\begin{array}{ll}
M, & \mathrm{for}~r\leq r_c,\\
M+\frac{2\ma}{3cr_g^{1/2}}\left(r^{3/2}-r_c^{3/2}\right), & \mathrm{for}~r>r_c.
\end{array}\right.
\end{equation}
Outside the sphere of critical radius, $r>r_c$, the above metric coincides with the case of $w=-0.5$ in Ref. \cite{Zeng:2020vsj}. However, one should be cautious that the physical interpretations are completely different \cite{Kiselev:2002dx,Visser:2019brz}, and here we are studying a different region of parameter space to estimate the gravitational effect of plasma particles.

In following sections, we will focus on two special models. All of the above equations are assumed to hold for both models. Their difference is merely the value of $r_c$, which is specified to $r_g$ in model A and to $3r_g$ in model B. Let us study them separately in Secs. \ref{sect-modA} and \ref{sect-modB}.

\section{Model A: $r_c=r_g$}\label{sect-modA}
In Ref. \cite{Perlick:2015vta}, it was implicitly assumed that the plasma density does not vanish on the photon sphere, which implies $r_c<3r_g/2$. In model A, we set $r_c=r_g$ to meet this condition concretely. Then Eq. \eqref{m} becomes
\begin{equation}\label{mA}
m(r)=\left\{\begin{array}{ll}
M, & \mathrm{for}~r\leq r_g,\\
M+\frac{2\ma}{3cr_g^{1/2}}\left(r^{3/2}-r_g^{3/2}\right), & \mathrm{for}~r>r_g.
\end{array}\right.
\end{equation}
Corresponding to \eqref{mA}, the spacetime has two horizons. One is the event horizon of black hole located at $r=r_g$, the other is the cosmological horizon at
\begin{equation}\label{chA}
r=\frac{r_g(1-\gamma)}{4\gamma^2}\left(\sqrt{1-\gamma}+\sqrt{1+3\gamma}\right)^2
\end{equation}
with a small parameter $\gamma$ defined in Eq. \eqref{gamma}. Both the photon sphere and the static observer live in the patch of spacetime between the two horizons.

\subsection{Influences on the radius}\label{subsect-radA}
With the model specified, we are now ready to compute the plasma's refractive and gravitational effects on the shadow accurately. Before entering on computations, let us make a rough comparison of the two effects. The refractive effect is induced intrinsically by the electromagnetic interaction. It is textbook knowledge that in a hydrogen atom, the ratio of the gravitational force to the electromagnetic force is $Gm_em_p/e^2=4.4\times10^{-40}$. Therefore, one would naively expect that the gravitational effect is suppressed by this factor in comparison with the refractive effect.

Let us take a closer look at model A. For this model, we can express the function $h(r)^2$ in terms of $\beta$ and $\gamma$ explicitly,
\begin{equation}\label{htoyA}
h(r)^2=r^2\left[1-\frac{r_g}{r}-\frac{\gamma}{r_g^{1/2}r}\left(r^{3/2}-r_g^{3/2}\right)\right]^{-1}-r^2\beta\left(\frac{r_g}{r}\right)^{3/2}.
\end{equation}
According to Eqs. \eqref{rph} and \eqref{bsh}, radii of the photon sphere and the shadow are determined simply by this function. It is clear that the refractive effect is controlled by $\beta$, while the gravitational effect is dictated by $\gamma$. For instance, one can switch off the gravitational effect by setting $\gamma=0$, then the above expression will reduce to Eq. (50) in Ref. \cite{Perlick:2015vta}. From this point of view, one would expect that the gravitational effect is suppressed by a factor $\gamma/\beta\sim Gm_em_pr_g^2/(e^2\lambda_0^2)$ in comparison with the refractive effect. Intriguingly, here is an enhancement factor $r_g^2/\lambda_0^2$ in competition with $Gm_em_p/e^2$.

As a further step, one can substitute Eq. \eqref{htoyA} into Eq. \eqref{rph} and search for the largest root, but the resulted equation cannot be solved exactly. To get some analytical results, we solve it perturbatively for small $\beta$ and $\gamma$. In this way, we find the second-order solution
\begin{equation}
\rph\approx r_{00}+r_{10}-r_{01}+r_{20}-r_{11}-r_{02},
\end{equation}
where
\begin{align}
\label{r01}&r_{00}=\frac{3}{2}r_g,~~~~r_{10}=\frac{\sqrt{6}}{108}\beta r_g,~~~~r_{01}=\left(\frac{3}{2}-\frac{9\sqrt{6}}{16}\right)\gamma r_g,\\
\label{r2}&r_{20}=\frac{7}{5832}\beta^2r_g,~~~~r_{11}=\left(\frac{1}{16}-\frac{\sqrt{6}}{216}\right)\beta\gamma r_g,~~~~r_{02}=\left(\frac{27\sqrt{6}}{32}-\frac{243}{128}\right)\gamma^2r_g.
\end{align}
With the help of Eq. \eqref{bsh}, we can proceed to compute the radius of shadow observed at a distance $\ro$. To the first order, the result is
\begin{equation}
\Rsh\approx R_{00}-R_{10}-R_{01}
\end{equation}
in which
\begin{align}
\label{b00A}&R_{00}=\frac{3\sqrt{3}}{2}r_g\left(1-\frac{r_g}{\ro}\right)^{1/2},\\
\label{b10A}&R_{10}=\left[\frac{\sqrt{2}}{6}-\frac{3\sqrt{3}}{4}\left(\frac{r_g}{\ro}\right)^{3/2}\left(1-\frac{r_g}{\ro}\right)\right]\beta r_g\left(1-\frac{r_g}{\ro}\right)^{1/2},\\
\label{b01A}&R_{01}=\left[\frac{3\sqrt{3}}{4}\left(\frac{\ro}{r_g}\right)^{1/2}+\frac{3}{8}\left(4\sqrt{3}-9\sqrt{2}\right)+\frac{3\sqrt{3}r_g^{1/2}}{4\left(\ro^{1/2}+r_g^{1/2}\right)}\right]\gamma r_g\left(1-\frac{r_g}{\ro}\right)^{1/2}.
\end{align}
It is straightforward to derive higher order terms. They are very lengthy but negligibly small, and thus not shown here. Remarkably, the gravitational correction to the radius of the black hole shadow is enhanced further by $(\ro/r_g)^{1/2}$ in Eq. \eqref{b01A}. In summary, the ratios of the gravitational corrections to the refractive corrections are
\begin{align}
\label{rphratA}&\frac{r_{01}}{r_{10}}\approx36\pi^2\left(4\sqrt{6}-9\right)\times\frac{Gm_em_pr_g^2}{e^2\lambda_0^2}\approx1.1\times10^{-6}\times\left(\frac{0.1\cm}{\lambda_0}\right)^2\left(\frac{M}{10^9\sm}\right)^2,\\
\label{RshratA}&\frac{R_{01}}{R_{10}}\approx12\sqrt{6}\pi^2\times\frac{Gm_em_pr_g^{3/2}\ro^{1/2}}{e^2\lambda_0^2}\approx0.4\times\left(\frac{0.1\cm}{\lambda_0}\right)^2\left(\frac{M}{10^9\sm}\right)^{3/2}\left(\frac{\ro}{10\mpc}\right)^{1/2}.
\end{align}

\begin{table}[ht]
\caption{Masses \cite{Firouzjaee:2019aij}, distances and mass accretion rates \cite{Daly:2020sex} for a sample of AGNs.}\label{tab-AGN}
\begin{tabular*}{\textwidth}{@{\extracolsep{\fill}}lccc}\hline
Source Name & Distance (Mpc) & Mass ($\sm$) & $\mathrm{Log}\left(dM/dt\right)$ ($\sm\mathrm{yr}^{-1}$) \\ \hline
Cyg A & $240$ & $1\times10^9$ & $-0.09$\\
NGC1275 & $75$ & $3.4\times10^8$ & $-0.87$\\
NGC3227 & $22$ & $4.22\times10^7$ & $-2.07$\\
NGC4151 & $22$ & $4\times10^7$ & $-1.69$\\
NGC4261 & $32$ & $4\times10^8$ & $-1.74$\\
NGC4374 & $20$ & $1.5\times10^9$ & $-1.80$\\
NGC4486 & $17$ & $6.77\times10^9$ & $-1.55$\\
NGC4594 & $11$ & $1\times10^9$ & $-1.75$\\
NGC5548 & $75$ & $6.71\times10^7$ & $-1.17$\\
NGC6166 & $124$ & $3\times10^{10}$ & $-1.71$\\
NGC7469 & $71$ & $1.22\times10^7$ & $-1.99$\\
Sgr A* & $0.008$ & $4.3\times10^6$ & $-6.23$\\
\hline\end{tabular*}
\end{table}

\begin{figure}
\centering
\includegraphics[width=0.45\textwidth]{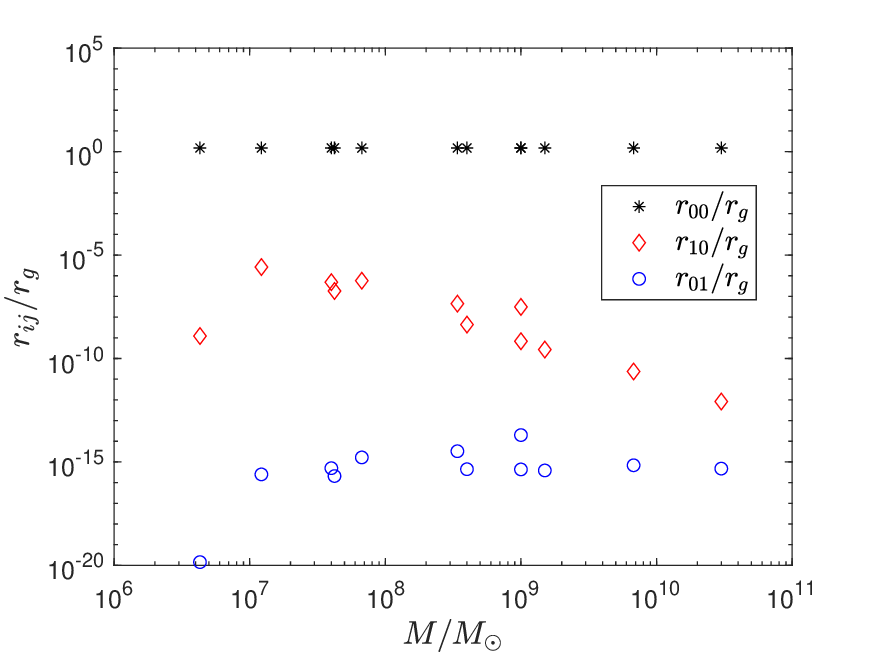}\includegraphics[width=0.45\textwidth]{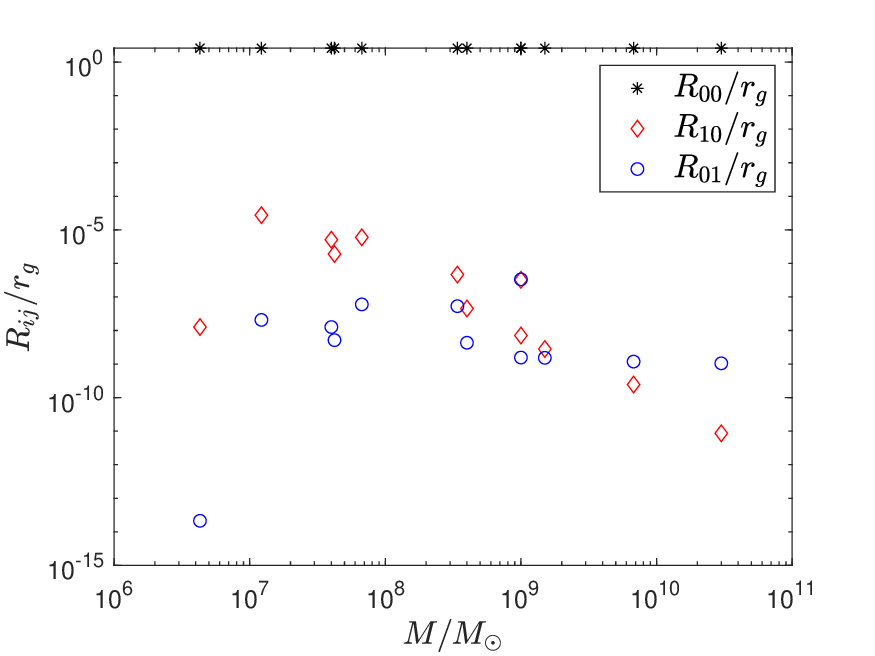}\\
\caption{(color online). Influences of the plasma on the radius of the black hole shadow for model A in Sec. \ref{sect-modA}. Left panel: the radius of the photon sphere and corrections to it according to Eq. \eqref{r01} and Table \ref{tab-AGN} with $\lambda_0=0.13\cm$. Right panel: the radius of the shadow and corrections to it according to Eqs. \eqref{b00A}, \eqref{b10A}, \eqref{b01A} and Table \ref{tab-AGN}. In both panels, black asterisks represent the uncorrected radii, red diamonds denote the refractive corrections, and  blue circles mark the gravitational corrections.}\label{fig-radA}
\end{figure}

To convert the above analytical expressions to numbers, we need the values of $\lambda_0$, $\ro$, $M$ and $\ma$. Table \ref{tab-AGN} is a sample of AGNs selected from Refs. \cite{Firouzjaee:2019aij,Daly:2020sex}. The data of masses are extracted from Ref. \cite{Firouzjaee:2019aij}, while the data of distances and mass accretion rates are obtained from Ref. \cite{Daly:2020sex}. Making use of the data and ignoring the spins of black holes, we evaluate Eq. \eqref{r01} for each AGN and draw the results in the left panel of Fig. \ref{fig-radA} in logarithmic coordinates, and Eqs. \eqref{b00A}, \eqref{b10A}, \eqref{b01A} in the right panel. In both panels, the uncorrected radii are drawn as asterisks, the refractive corrections are depicted by diamonds, and the gravitational corrections are denoted by circles. For all of them, we have set $\lambda_0=0.13\cm$ \cite{Akiyama:2019cqa}. This figure can be regarded as a semi-realistic estimation of refractive and gravitational effects of plasma on the radii of the photon sphere and the shadow. We can see clearly that these effects are negligible despite of the diversity of AGNs in the sample. The right panel is in good agreement with Eq. \eqref{RshratA}, which shows that the two effects on the radius of shadow are comparable for black holes of mass $M\sim10^9\sm$. For Sgr A*, the gravitational corrections are smallest because of its slowest accretion rate.

\subsection{Influences on the intensity}\label{sect-intA}
From the conclusion of the previous subsection, one can infer that the refractive and gravitational effects on the intensity image should be undetectable for realistic AGNs. This is due to the smallness of $\beta$ and $\gamma$ in Eqs. \eqref{beta}, \eqref{gamma}. In the current subsection, to make the effects noticeable, we will assign unrealistically large values to these parameters.

\begin{figure}
\centering
\includegraphics[width=0.45\textwidth]{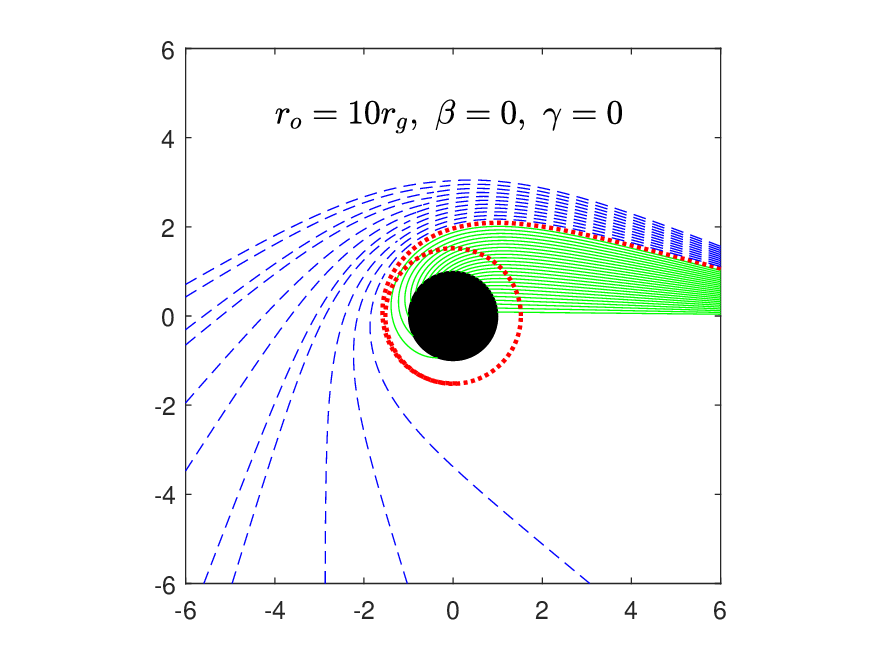}\includegraphics[width=0.45\textwidth]{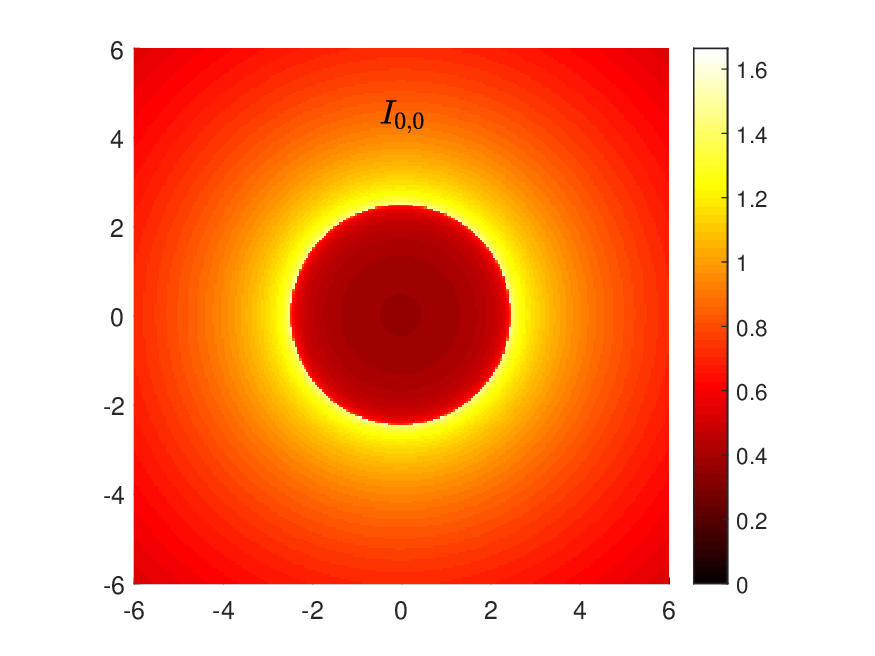}\\
\includegraphics[width=0.45\textwidth]{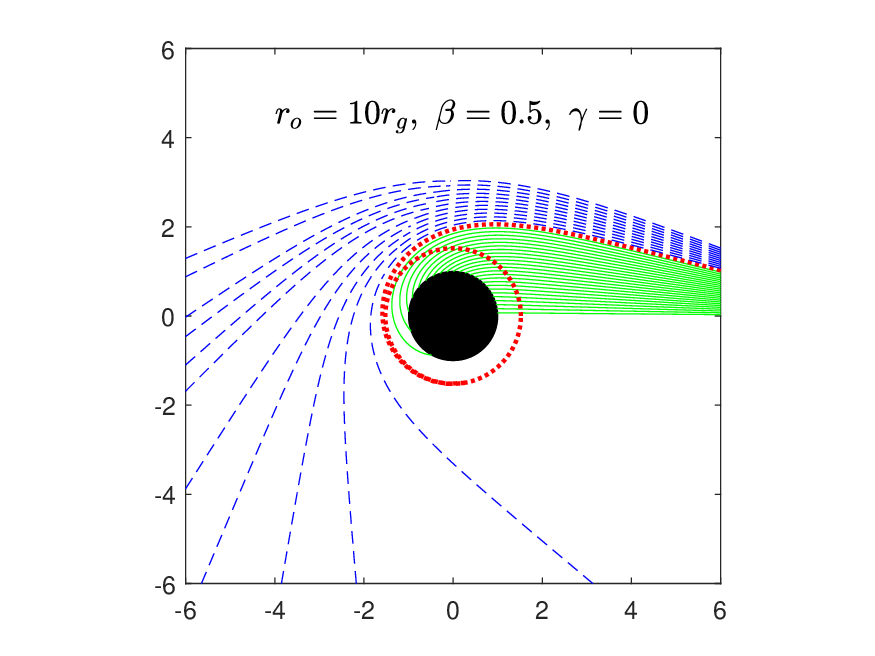}\includegraphics[width=0.45\textwidth]{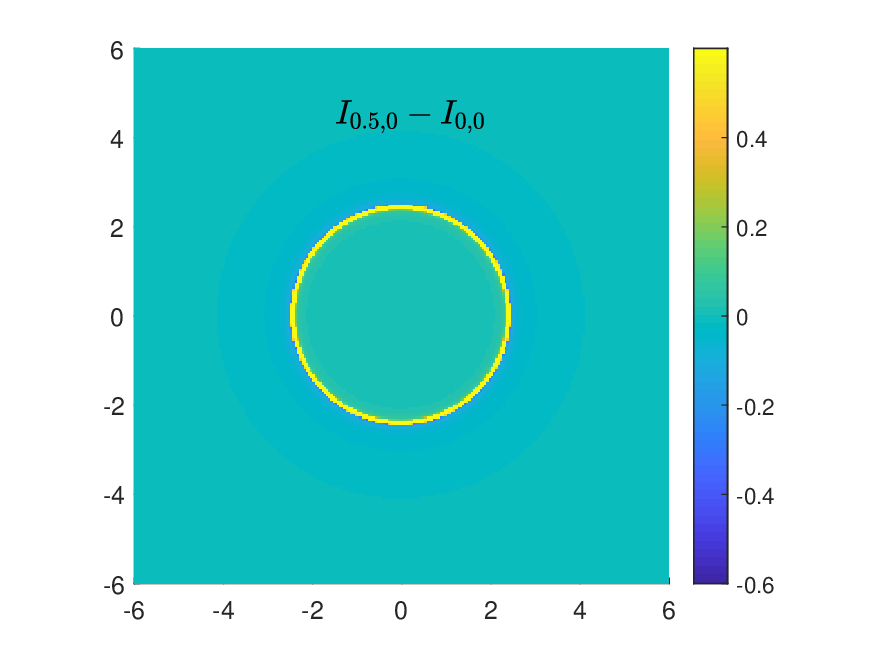}\\
\includegraphics[width=0.45\textwidth]{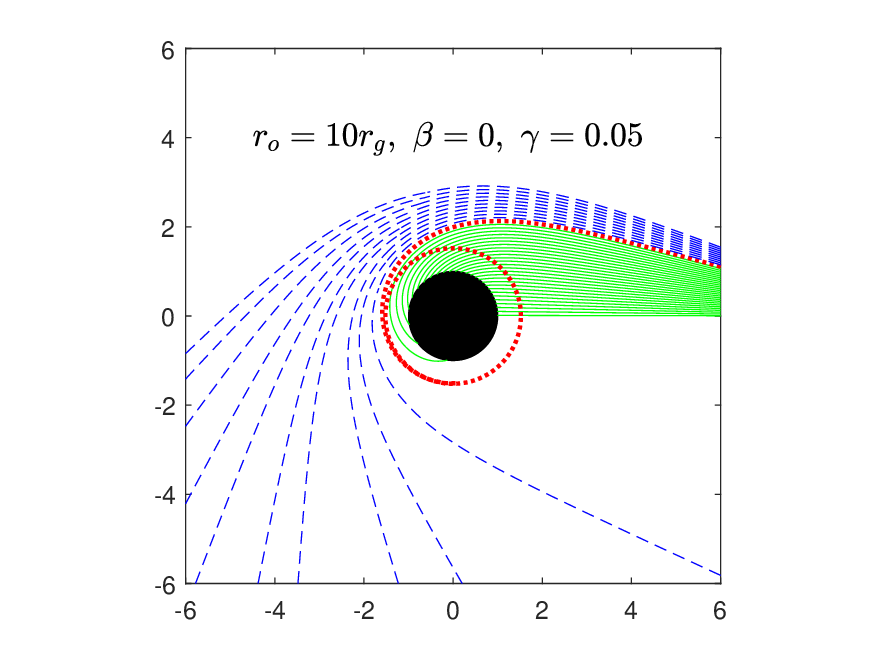}\includegraphics[width=0.45\textwidth]{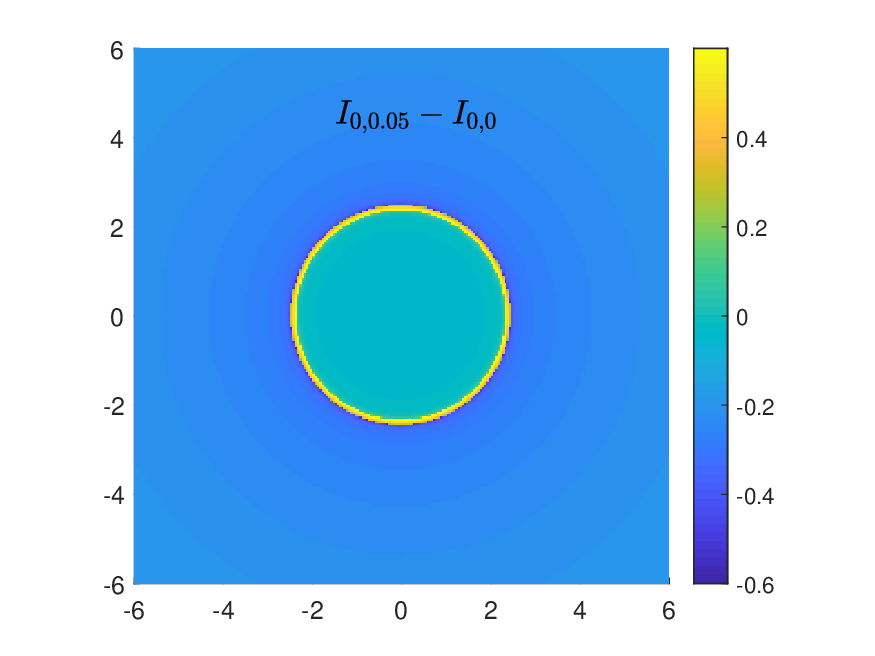}\\
\caption{(color online). Influences of the plasma on the intensity of the black hole shadow for model A in Sec. \ref{sect-modA}. Left panels: trajectories of the light rays observed at the point $(x,y)=(10r_g,0)$. Green solid curves, red dotted curves and blue dashed curves depict trajectories with $b<h(\rph)$, $b=h(\rph)$ and $b>h(\rph)$ respectively. The interval of adjacent orbits is $\Delta b/r_g=0.1$. From top left to bottom left, the critical trajectories $b=h(\rph)$ correspond to $\rph/r_g=1.5,1.511647,1.493443$ and $h(\rph)/r_g=2.598076,2.477186,2.714083$ in turn. Right panels: the observed intensity image for $\ro=10r_g$, $\beta=0$, $\gamma=0$ (top right) and its differences with images for $\beta=0.5$, $\gamma=0$ (middle right) and $\beta=0$, $\gamma=0.05$ (bottom right). All Axes are rescaled by $r_g$.}\label{fig-intA}
\end{figure}

For the toy model A, the orbit equation \eqref{orbitr} of light rays can be reformed as
\begin{equation}
\frac{r_g}{D(r)}\frac{dr}{d\phi}=\pm\sqrt{\frac{r_g^2}{b^2}\left[1-A\frac{\omp(r)^2}{\omega_0^2}\right]-\frac{r_g^2A}{D(r)}}.
\end{equation}
In terms of a new variable $u=r_g^{1/2}/r^{1/2}$, we can write down $\omp(r)^2/\omega_0^2=\beta u^3$, $D(r)=r_g^2u^{-4}$ and
\begin{equation}\label{AuA}
A=u^{-1}\left[u-(1-\gamma)u^3-\gamma\right]
\end{equation}
outside the event horizon. For clockwise light rays $du/d\phi>0$, the orbit equation can be reexpressed in terms of $u$ as
\begin{equation}\label{orbituA}
2u\frac{du}{d\phi}=\sqrt{\frac{r_g^2}{b^2}\left(1-\beta u^3A\right)-u^4A}.
\end{equation}
When performing the numerical integration, exaggerated values are assigned to the model parameters as labeled in Fig. \ref{fig-intA}, $\ro=10r_g$, $\beta=0$ or $0.5$, $\gamma=0$ or $0.05$.

Let us consider an observer located outside the photon sphere. According to the ratio $b/h(\rph)$, the light rays arriving at this observer can be classified into three types \cite{Luminet:1979nyg} which are depicted by different line-types in Fig. \ref{fig-intA}: (i) Light rays with $b<h(\rph)$ can travel through the photon sphere. As illustrated by green solid curves in Fig. \ref{fig-intA}, all orbits of these rays start near the event horizon of the black hole. (ii) A critical light ray has $b=h(\rph)$. It propagates in an unstable circular orbit of radius $\rph$ and has a chance to escape to the observer under radial perturbations. The circle is a great circle of the photon sphere. In every left panel of Fig. \ref{fig-intA}, we plot such a critical light ray as a red dotted curve. (iii) Light rays with $b>h(\rph)$ are always outside the photon sphere. As depicted by blue dashed curves in Fig. \ref{fig-intA}, each orbit of such rays is symmetric with respect to a diametrical line through its pericenter. The radial coordinate $\rmin$ of the pericenter is dictated by $h(\rmin)=b$.

It is reasonable to expect that the radiant energy density is proportional to the plasma density. This implies a specific emissivity in the rest-frame of the emitter
\begin{equation}\label{jA}
j(\nu,r)\propto\frac{1}{r^{3/2}}.
\end{equation}
For simplicity we assume the emission is uniform in frequency from a static source. Then $r^{3/2}j(\nu,r)$ is independent of $\nu$ and $r$, and thus the specific intensity at the point $(x,y)$ of the observed image is \cite{Jaroszynski:1997bw,Bambi:2013nla,Zeng:2020vsj}
\begin{equation}
I_{\beta,\gamma}=\frac{1}{2}\mathcal{I}r_g^{1/2}\int_{\mathrm{ray}}\frac{A(r)^{3/2}}{r^{3/2}}\sqrt{B(r)+r^2\left(\frac{d\phi}{dr}\right)^2}dr.
\end{equation}
Here the normalization $\mathcal{I}$ is unimportant in our simulations, and the subscripts are used to reminding us the dependence of intensity on $\beta$ and $\gamma$. Numerically it is more convenient to perform this integration in terms of $u$ defined above,
\begin{equation}\label{IA}
I_{\beta,\gamma}=\mathcal{I}\int_{\mathrm{ray}}A^{3/2}\sqrt{\frac{1}{A}+\frac{u^2}{4}\left(\frac{d\phi}{du}\right)^2}du.
\end{equation}
The integration is performed using the backward ray shooting method \cite{Luminet:1979nyg,Jaroszynski:1997bw}. For type (i) light rays, we integrate Eq. \eqref{IA} from the observer to the event horizon of the black hole. For type (iii) light rays, the integration is performed from the observer to the pericenter and then to the cosmological horizon \eqref{chA}. In practice, telescopes do not collect light rays with $\pi/2<\alpha\leq\pi$, thus we assume $0\leq\alpha\leq\pi/2$ in our simulations.

By virtue of the spherical symmetry of the toy model, the image is rotationally invariant, and thus the intensity is dependent simply on $R=\left(x^2+y^2\right)^{1/2}$. By definition Eq. \eqref{R}, it is easy to show
\begin{equation}\label{Ru}
R^2=\frac{b^2A}{1-\beta u^3A}
\end{equation}
with $u=r_g^{1/2}/\ro^{1/2}$. By continuously changing the value of $b$, we have computed the dependence of intensity on $R$ and simulated the images with $\ro=10r_g$, $\beta=0$ or $0.5$, $\gamma=0$ or $0.05$. The slight differences between these images are not easy to notice, especially near the ring. In the top right panel of Fig. \ref{fig-intA}, we present the image for $I_{0,0}$. In the middle right and bottom right panels, we subtract $I_{0,0}$ from $I_{0.5,0}$ and $I_{0,0.05}$ to illustrate the slight differences. In comparison with the case of $\beta=0,\gamma=0$, the intensity outside the ring decreases more significantly in the case of $\beta=0,\gamma=0.05$ than the case of $\beta=0.5,\gamma=0$, but the changes in radius are about the same.

\section{Model B: $r_c=3r_g$}\label{sect-modB}
In Ref. \cite{1972ApSS15153M}, dropping corrections to the metric from the external plasma, it was shown that no physically acceptable solution exists if $r_c<3r_g$. In other words, the critical radius ought to be slightly larger than the radius of the innermost stable circular orbit. In model B, we consider the limiting case and set $r_c=3r_g$. Then Eq. \eqref{m} becomes
\begin{equation}\label{mB}
m(r)=\left\{\begin{array}{ll}
M, & \mathrm{for}~r\leq3r_g,\\
M+\frac{2\ma}{3cr_g^{1/2}}\left(r^{3/2}-3\sqrt{3}r_g^{3/2}\right), & \mathrm{for}~r>3r_g.
\end{array}\right.
\end{equation}
Corresponding to \eqref{mB}, the event horizon of black hole is located at $r=r_g$, and the cosmological horizon is located at
\begin{equation}\label{chB}
r=\frac{r_g}{9\gamma^2}\left[1+\left(C+\sqrt{C^2-1}\right)^{1/3}+\left(C-\sqrt{C^2-1}\right)^{1/3}\right]^2
\end{equation}
with
\begin{equation}
C=\frac{1}{2}\left(2-27\gamma^2+81\sqrt{3}\gamma^3\right).
\end{equation}
Both the photon sphere and the static observer live in the patch of spacetime between the two horizons.

\subsection{Influences on the radius}\label{subsect-radB}
In model A, the density profile of the plasma is a continuous function outside the event horizon. In model B, however, the density profile has a step at the innermost stable circular orbit. Thereby the function $h(r)^2$ is discontinuous, taking the form
\begin{equation}\label{htoyB}
h(r)^2=\left\{\begin{array}{ll}
r^2\left(1-\frac{r_g}{r}\right)^{-1}, & \mathrm{for}~r\leq3r_g,\\
r^2\left[1-\frac{r_g}{r}-\frac{\gamma}{r_g^{1/2}r}\left(r^{3/2}-3\sqrt{3}r_g^{3/2}\right)\right]^{-1}-r^2\beta\left(\frac{r_g}{r}\right)^{3/2}, & \mathrm{for}~r>3r_g.
\end{array}\right.
\end{equation}
Applying Eq. \eqref{rph} to this function, we find the radius of the photon sphere can be derived from the function in the region $r\leq3r_g$ as long as $\beta$ and $\gamma$ are small. The result is
\begin{equation}
\rph=\frac{3}{2}r_g,
\end{equation}
the same as the Schwarzschild black hole without a plasma. Indeed, Eq. \eqref{rph} indicates that $\rph$ is determined locally by $h(r)^2$ near the photon sphere as long as the equation has no root outside the sphere. Reversing this logic, one can infer that the plasma density does not vanish on the photon sphere in Ref. \cite{Perlick:2015vta}, because $\rph^1\neq0$ in Eq. (53) of Ref. \cite{Perlick:2015vta}.

Although the value of $\rph$ is trivial in model B, the value of $\Rsh$ is nontrivial. According to Eq. \eqref{bsh}, the radius of the shadow depends on both the radius of the photon sphere and the radial coordinate of the observer. As an analytical result, the radius of shadow observed at a distance $\ro$ is
\begin{equation}
\Rsh\approx R_{00}+R_{10}-R_{01}
\end{equation}
to the first order of $\beta$ and $\gamma$, in which
\begin{align}
\label{b00B}&R_{00}=\frac{3\sqrt{3}}{2}r_g\left(1-\frac{r_g}{\ro}\right)^{1/2},\\
\label{b10B}&R_{10}=\frac{3\sqrt{3}}{4}\left(\frac{r_g}{\ro}\right)^{3/2}\left(1-\frac{r_g}{\ro}\right)^{3/2}\beta r_g,\\
\label{b01B}&R_{01}=\left[\frac{3\sqrt{3}}{4}\left(\frac{\ro}{r_g}\right)^{1/2}+\frac{3\sqrt{3}r_g^{1/2}}{4\left(\ro^{1/2}+r_g^{1/2}\right)}\right]\gamma r_g\left(1-\frac{r_g}{\ro}\right)^{1/2}.
\end{align}
In this model, the ratio of the gravitational correction to the refractive correction is
\begin{equation}\label{RshratB}
\frac{R_{01}}{R_{10}}\approx\frac{16\pi^2}{3}\times\frac{Gm_em_p\ro^2}{e^2\lambda_0^2}\approx2\times10^{15}\times\left(\frac{0.1\cm}{\lambda_0}\right)^2\left(\frac{\ro}{10\mpc}\right)^2.
\end{equation}

\begin{figure}
\centering
\includegraphics[width=0.45\textwidth]{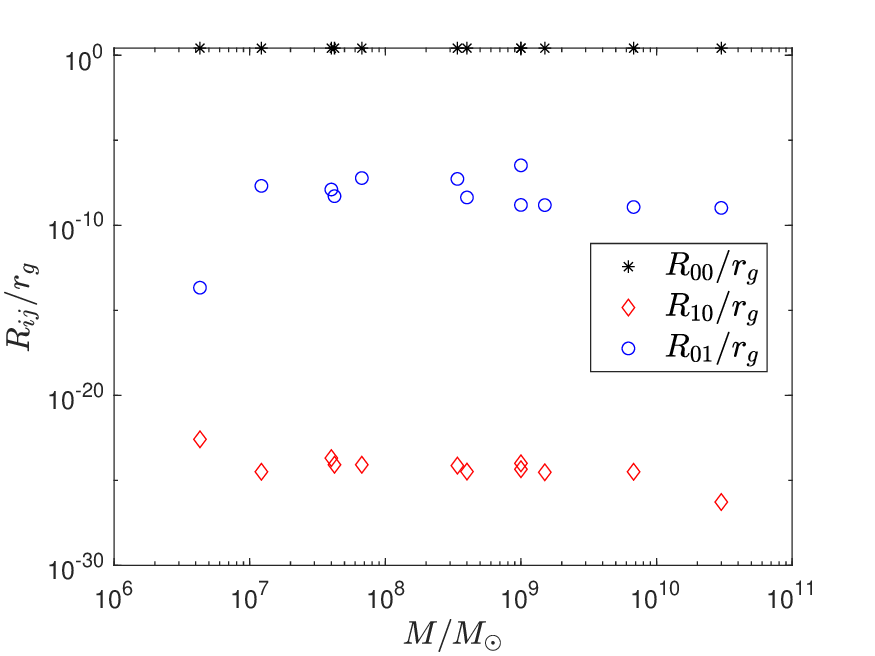}\\
\caption{(color online). Influences of the plasma on the radius of the black hole shadow for model B in Sec. \ref{sect-modB}. Black asterisks represent the uncorrected radius according to Eq. \eqref{b00B}, red diamonds denote the refractive corrections according to Eq. \eqref{b10B}, and  blue circles mark the gravitational corrections according to Eq. \eqref{b01B}. Data in Table \ref{tab-AGN} have been used here.}\label{fig-radB}
\end{figure}

Making use of the data in Sec. \ref{subsect-radA}, we evaluate Eqs. \eqref{b00B}, \eqref{b10B}, \eqref{b01B} for each AGN and draw the results in Fig. \ref{fig-radB}.

\subsection{Influences on the intensity}\label{sect-intB}
\begin{figure}
\centering
\includegraphics[width=0.45\textwidth]{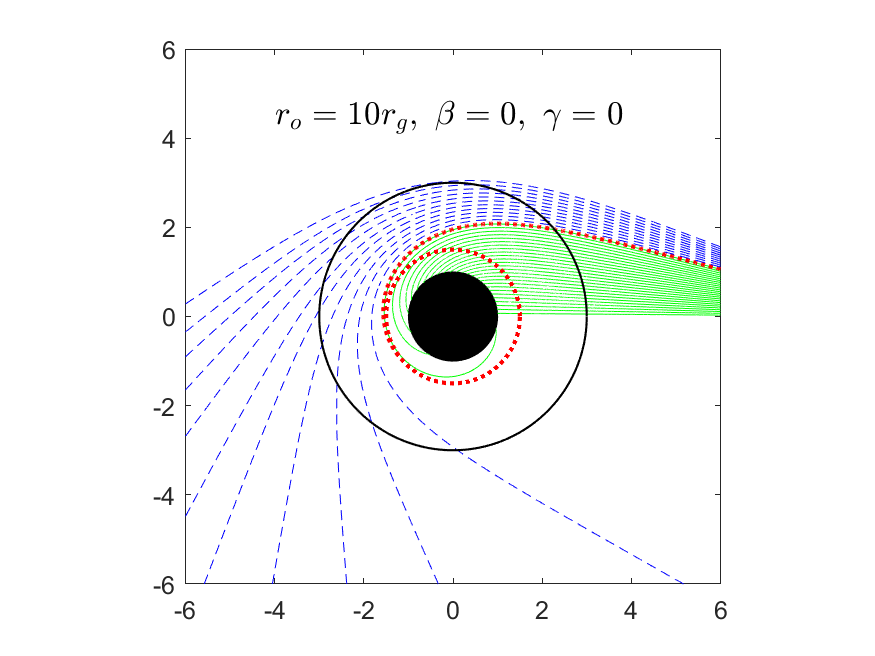}\includegraphics[width=0.45\textwidth]{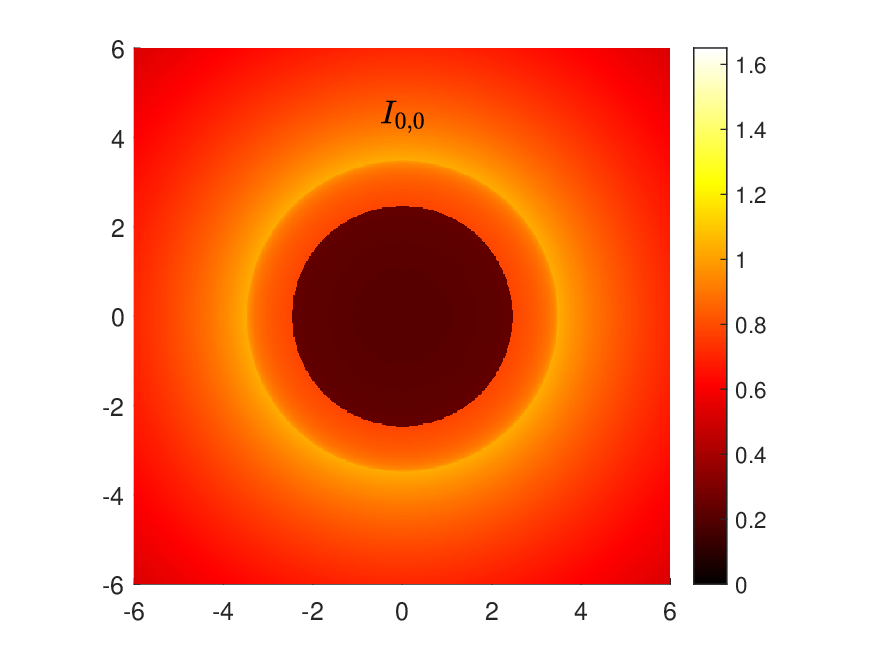}\\
\includegraphics[width=0.45\textwidth]{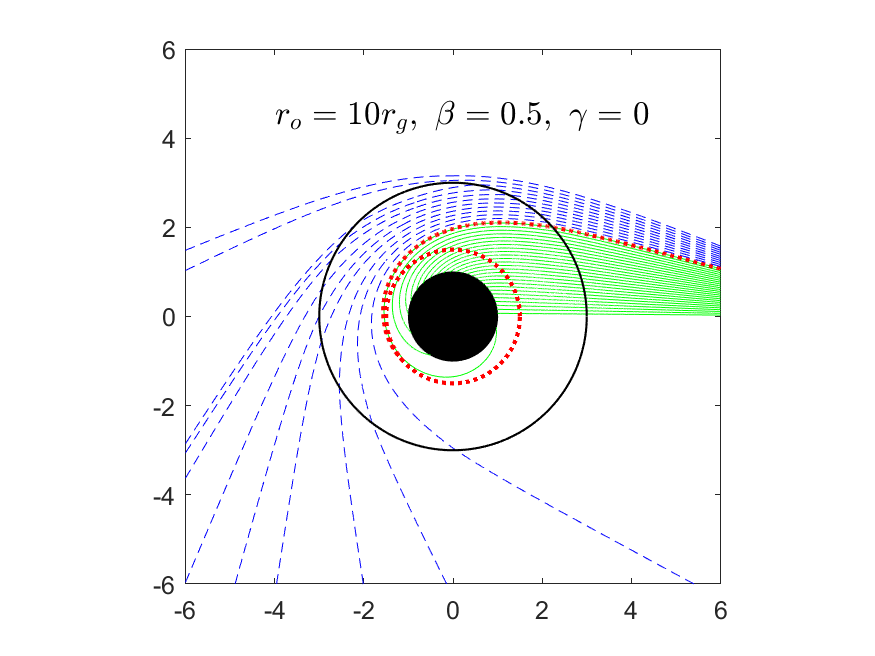}\includegraphics[width=0.45\textwidth]{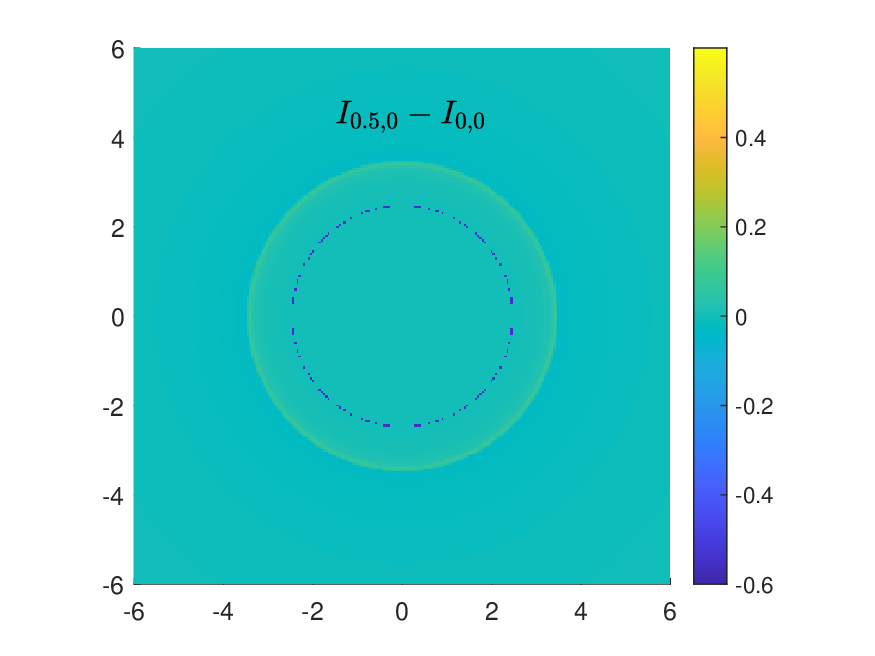}\\
\includegraphics[width=0.45\textwidth]{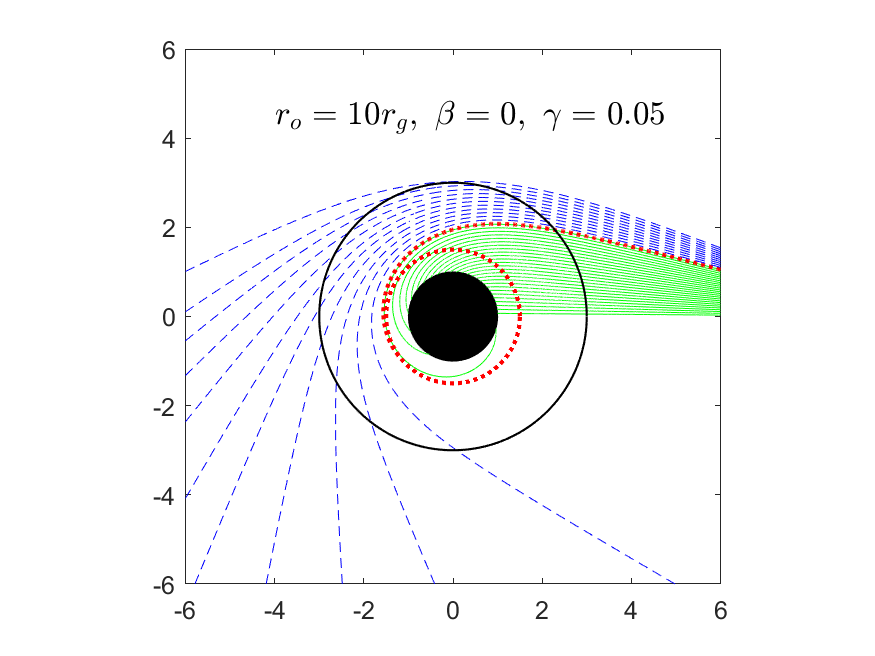}\includegraphics[width=0.45\textwidth]{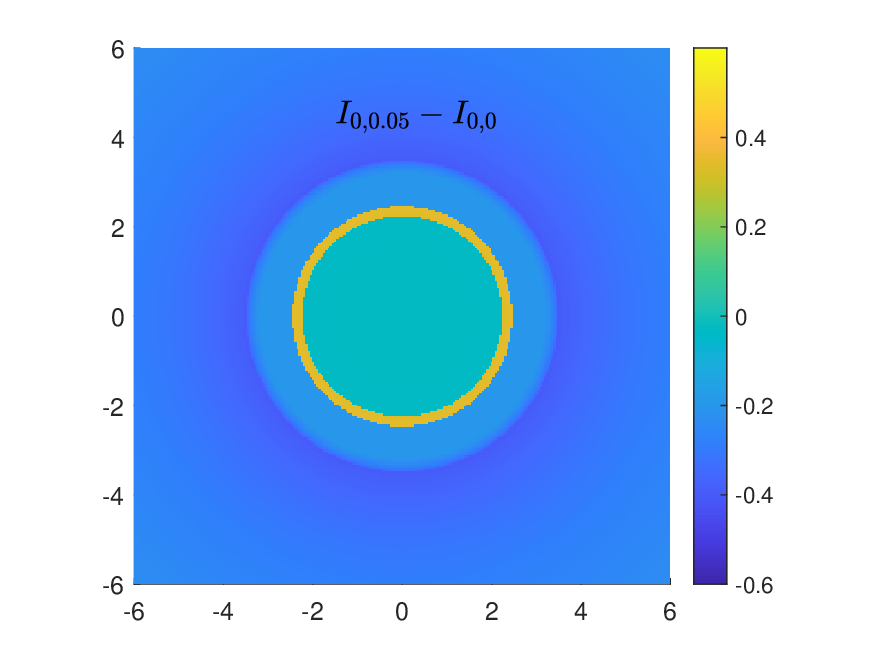}\\
\caption{(color online). Influences of the plasma on the intensity of the black hole shadow for model B in Sec. \ref{sect-modB}. Left panels: trajectories of the light rays observed at the point $(x,y)=(10r_g,0)$. Green solid curves, red dotted curves and blue dashed curves depict trajectories with $b<h(\rph)$, $b=h(\rph)$ and $b>h(\rph)$ respectively. The interval of adjacent orbits is $\Delta b/r_g=0.1$. Right panels: the observed intensity image for $\ro=10r_g$, $\beta=0$, $\gamma=0$ (top right) and its differences with images for $\beta=0.5$, $\gamma=0$ (middle right) and $\beta=0$, $\gamma=0.05$ (bottom right). All Axes are rescaled by $r_g$.}\label{fig-intB}
\end{figure}

In terms of $u=r_g^{1/2}/r^{1/2}$, the orbit equation for clockwise light rays can be expressed as
\begin{equation}\label{orbituB}
2u\frac{du}{d\phi}=\left\{\begin{array}{ll}
\sqrt{\frac{r_g^2}{b^2}\left(1-\beta u^3A\right)-u^4A}, & \mathrm{for}~u<1/\sqrt{3},\\
\sqrt{\frac{r_g^2}{b^2}-u^4\left(1-u^2\right)}, & \mathrm{for}~1/\sqrt{3}\leq u\leq1
\end{array}\right.
\end{equation}
in model B, where
\begin{equation}\label{AuB}
A=u^{-1}\left[u-(1-3\sqrt{3}\gamma)u^3-\gamma\right].
\end{equation}
Considering an observer located at $\ro=10r_g$, we illustrate the trajectories of light rays in left panels of Fig.\ref{fig-intB} for $\beta=0$ or $0.5$, $\gamma=0$ or $0.05$. The innermost stable circular orbit is depicted by a big black circle. Noticeably, when $\beta\neq0$, the behaviors of light rays in the neighborhoods of the innermost stable circular orbit are distinct. This can be understood from
\begin{equation}
\lim_{u\rightarrow1/\sqrt{3}}\sqrt{\frac{r_g^2}{b^2}\left(1-\beta u^3A\right)-u^4A}=\sqrt{\frac{r_g^2}{b^2}\left(1-\frac{2\beta}{9\sqrt{3}}\right)-\frac{2}{27}},
\end{equation}
which suggests that the bending of light is weakened by the refractive effect near the innermost stable circular orbit.

In model B, the plasma is the same as in model A when $r>3r_g$ but vanishes when $r\leq3r_g$. Therefore, the specific emissivity has the form
\begin{equation}\label{jB}
j(\nu,r)\propto\frac{1}{r^{3/2}}H(r-3r_g).
\end{equation}
Here the Heaviside step function is defined as
\begin{equation}
H(x)=\left\{\begin{array}{ll}
0, & \mathrm{for}~x\leq0,\\
1, & \mathrm{for}~x>0.
\end{array}\right.
\end{equation}
When computing the specific intensity of the observed image in this model, we should insert the Heaviside unit step function into Eq. \eqref{IA},
\begin{equation}\label{IB}
I_{\beta,\gamma}=\mathcal{I}\int_{\mathrm{ray}}A^{3/2}\sqrt{\frac{1}{A}+\frac{u^2}{4}\left(\frac{d\phi}{du}\right)^2}H\left(\frac{1}{\sqrt{3}}-u\right)du
\end{equation}
and take $A$ as Eq. \eqref{AuB}. The images for model B are presented in right panels of Fig.\ref{fig-intB}. In this model, the image of photon sphere makes negligible contributions to the total brightness. Similar results have been reported previously in Refs. \cite{Zeng:2020vsj,Gralla:2019xty,He:2021htq}.

\section{Discussion}\label{sect-dis}
In this paper, we have investigated the gravitational effect of plasma particles on the radius and the intensity of the black hole shadow. We did this in a toy model of a Schwarzschild black hole surrounded by a static plasma of density $\rho(r)\propto r^{-3/2}$ outside the event horizon. Inserting observational data for a sample of AGNs into our analytical results, we find the gravitational correction $R_{01}$ to the shadow radius is in the range $10^{-8}\sim10^{-6}$. In contrast, we find the correction $R_{10}$ from the refractive effect of plasma medium decreases prominently as the black hole mass increases, covering the range $10^{-4}\sim10^{-10}$. Alternatively, turning to the scenario that the space is empty between event horizon and innermost stable circular orbit, we find the shadow is not determined by the photon sphere, but by the inner boundary of the accretion disk.

Strictly speaking, our analytical results are reliable only for the static toy model\cite{Claudel:2000yi}, so the numbers above are very rough estimations for realistic AGNs which usually have nonnegligible spins \cite{Wei:2019pjf,Banerjee:2019nnj}. However, as a concrete example, they demonstrate that the gravitational effect of plasma particles can exceed the refractive effect of the plasma medium under some circumstances. The gravitational effect is more complicated to compute than the refractive effect. If the refractive effect is proven to be detectable in a more realistic model, then the gravitational effect should be scrutinized, especially for AGNs of high masses.

We did not consider synchrotron self-absorption \cite{Falcke:1999pj}, which is desirable when studying the background emission of the plasma. It would also be interesting to take the black hole spin \cite{Atamurotov:2015nra,Perlick:2017fio,Huang:2018rfn,Yan:2019etp,Cunha:2019hzj,Chowdhuri:2020ipb,Konoplya:2021slg,Badia:2021kpk} and the QED effect \cite{Hu:2020usx} into account.

\bibliographystyle{utcaps}
\bibliography{Reference}

\end{document}